\documentclass[aps,prl,twocolumn,groupedaddress,superscriptaddress,nofootinbib,notitlepage,floatfix]{revtex4-2}
\usepackage{times,bm,bbm,amssymb,amsmath,amsfonts,graphics,graphicx,color,hyperref}
\hypersetup{colorlinks,linkcolor={blue},citecolor={blue},urlcolor= {blue}}
\usepackage{graphicx}
\usepackage{float}
\usepackage{siunitx}
\usepackage{amsmath}
\hypersetup{colorlinks,linkcolor={blue},citecolor={blue},urlcolor= {blue}}
\newcommand{\ignore}[1]{}
\DeclareFontFamily{OT1}{pzc}{}
\DeclareFontShape{OT1}{pzc}{m}{it}
              {<-> s * [1.25] pzcmi7t}{}
\DeclareMathAlphabet{\mathpzc}{OT1}{pzc}
                                 {m}{it}
\newcommand{\ket}[1]{\mathinner{|{#1}\rangle}}

\newcommand{\bra}[1]{\mathinner{\langle{#1}|}}

\newcommand{\Bra}[1]{\left<#1\right|}
\newcommand{\Ket}[1]{\left|#1\right >}      

\let\vec\oldvec
\newcommand{\vec}{\mathbf}
\newcommand{\mat}{\textsf}

\newcommand{\QCD}{\affiliation{QCD Labs, QTF Centre of Excellence, Department of Applied Physics, Aalto University, P.O.~Box 13500, FI-00076 Aalto, Finland}}

\newcommand{\VTT}{\affiliation{QTF Centre of Excellence, VTT Technical Research Centre of Finland Ltd, P.O. Box 1000, FI-02044 VTT, Finland}}

\newcommand{\Oulu}{\affiliation{Nano and Molecular Systems Research Unit, University of Oulu, P.O.~Box 3000, FI-90014 Oulu, Finland}}               
\usepackage[T1]{fontenc}
\usepackage{newtxtext,newtxmath}
\begin{document}
\author{Aravind P. Babu}
\Oulu
\author{Tuure Orell}
\Oulu
\author{Vasilii Vadimov}
\QCD
\author{Wallace Teixeira}
\QCD
\global\long\def\ha{\hat{a}}%

\global\long\def\hb{\hat{b}}%

\global\long\def\hc{\hat{c}}%

\global\long\def\hd{\hat{d}}%

\global\long\def\he{\hat{e}}%

\global\long\def\hf{\hat{f}}%

\global\long\def\hg{\hat{g}}%

\global\long\def\hh{\hat{h}}%

\global\long\def\hi{\hat{i}}%

\global\long\def\hj{\hat{j}}%

\global\long\def\hk{\hat{k}}%

\global\long\def\hl{\hat{l}}%

\global\long\def\hm{\hat{m}}%

\global\long\def\hn{\hat{n}}%

\global\long\def\ho{\hat{o}}%

\global\long\def\hp{\hat{p}}%

\global\long\def\hq{\hat{q}}%

\global\long\def\hr{\hat{r}}%

\global\long\def\hs{\hat{s}}%

\global\long\def\hu{\hat{u}}%

\global\long\def\hv{\hat{v}}%

\global\long\def\hw{\hat{w}}%

\global\long\def\hx{\hat{x}}%

\global\long\def\hy{\hat{y}}%

\global\long\def\hz{\hat{z}}%

\global\long\def\hA{\hat{A}}%

\global\long\def\hB{\hat{B}}%

\global\long\def\hC{\hat{C}}%

\global\long\def\hD{\hat{D}}%

\global\long\def\hE{\hat{E}}%

\global\long\def\hF{\hat{F}}%

\global\long\def\hG{\hat{G}}%

\global\long\def\hH{\hat{H}}%

\global\long\def\hI{\hat{I}}%

\global\long\def\hJ{\hat{J}}%

\global\long\def\hK{\hat{K}}%

\global\long\def\hL{\hat{L}}%

\global\long\def\hM{\hat{M}}%

\global\long\def\hN{\hat{N}}%

\global\long\def\hO{\hat{O}}%

\global\long\def\hP{\hat{P}}%

\global\long\def\hQ{\hat{Q}}%

\global\long\def\hR{\hat{R}}%

\global\long\def\hS{\hat{S}}%

\global\long\def\hT{\hat{T}}%

\global\long\def\hU{\hat{U}}%

\global\long\def\hV{\hat{V}}%

\global\long\def\hW{\hat{W}}%

\global\long\def\hX{\hat{X}}%

\global\long\def\hY{\hat{Y}}%

\global\long\def\hZ{\hat{Z}}%

\global\long\def\hap{\hat{\alpha}}%

\global\long\def\hbt{\hat{\beta}}%

\global\long\def\hgm{\hat{\gamma}}%

\global\long\def\hGm{\hat{\Gamma}}%

\global\long\def\hdt{\hat{\delta}}%

\global\long\def\hDt{\hat{\Delta}}%

\global\long\def\hep{\hat{\epsilon}}%

\global\long\def\hvep{\hat{\varepsilon}}%

\global\long\def\hzt{\hat{\zeta}}%

\global\long\def\het{\hat{\eta}}%

\global\long\def\hth{\hat{\theta}}%

\global\long\def\hvth{\hat{\vartheta}}%

\global\long\def\hTh{\hat{\Theta}}%

\global\long\def\hio{\hat{\iota}}%

\global\long\def\hkp{\hat{\kappa}}%

\global\long\def\hld{\hat{\lambda}}%

\global\long\def\hLd{\hat{\Lambda}}%

\global\long\def\hmu{\hat{\mu}}%

\global\long\def\hnu{\hat{\nu}}%

\global\long\def\hxi{\hat{\xi}}%

\global\long\def\hXi{\hat{\Xi}}%

\global\long\def\hpi{\hat{\pi}}%

\global\long\def\hPi{\hat{\Pi}}%

\global\long\def\hrh{\hat{\rho}}%

\global\long\def\hvrh{\hat{\varrho}}%

\global\long\def\hsg{\hat{\sigma}}%

\global\long\def\hSg{\hat{\Sigma}}%

\global\long\def\hta{\hat{\tau}}%

\global\long\def\hup{\hat{\upsilon}}%

\global\long\def\hUp{\hat{\Upsilon}}%

\global\long\def\hph{\hat{\phi}}%

\global\long\def\hvph{\hat{\varphi}}%

\global\long\def\hPh{\hat{\Phi}}%

\global\long\def\hch{\hat{\chi}}%

\global\long\def\hps{\hat{\psi}}%

\global\long\def\hPs{\hat{\Psi}}%

\global\long\def\hom{\hat{\omega}}%

\global\long\def\hOm{\hat{\Omega}}%

\global\long\def\hdgg#1{\hat{#1}^{\dagger}}%

\global\long\def\cjg#1{#1^{*}}%

\global\long\def\hsgx{\hat{\sigma}_{x}}%

\global\long\def\hsgy{\hat{\sigma}_{y}}%

\global\long\def\hsgz{\hat{\sigma}_{z}}%

\global\long\def\hsgp{\hat{\sigma}_{+}}%

\global\long\def\hsgm{\hat{\sigma}_{-}}%

\global\long\def\hsgpm{\hat{\sigma}_{\pm}}%

\global\long\def\hsgmp{\hat{\sigma}_{\mp}}%

\global\long\def\dert#1{\frac{d}{dt}#1}%

\global\long\def\dertt#1{\frac{d#1}{dt}}%

\global\long\def\Tr{\text{Tr}}%


\global\long\def\ket#1{|#1\rangle}%

\global\long\def\Ket#1{\left|#1\right>}%

\global\long\def\bra#1{\langle#1|}%

\global\long\def\Bra#1{\left<#1\right|}%

\global\long\def\bk#1#2{\langle#1|#2\rangle}%

\global\long\def\BK#1#2{\left\langle #1\middle|#2\right\rangle }%

\global\long\def\kb#1#2{\ket{#1}\!\bra{#2}}%

\global\long\def\KB#1#2{\Ket{#1}\!\Bra{#2}}%

\global\long\def\mel#1#2#3{\bra{#1}#2\ket{#3}}%

\global\long\def\MEL#1#2#3{\Bra{#1}#2\Ket{#3}}%

\global\long\def\n#1{|#1|}%

\global\long\def\N#1{\left|#1\right|}%

\global\long\def\ns#1{|#1|^{2}}%

\global\long\def\NS#1{\left|#1\right|^{2}}%

\global\long\def\nn#1{\lVert#1\rVert}%

\global\long\def\NN#1{\left\lVert #1\right\rVert }%

\global\long\def\nns#1{\lVert#1\rVert^{2}}%

\global\long\def\NNS#1{\left\lVert #1\right\rVert ^{2}}%

\global\long\def\ev#1{\langle#1\rangle}%

\global\long\def\EV#1{\left\langle #1\right\rangle }%

\author{Mikko M\"{o}tt\"{o}nen}
\QCD
\VTT
\author{Matti Silveri}
\Oulu

\title{Quantum error correction under numerically exact open-quantum-system dynamics} 
\date{\today}
\begin{abstract}
The known quantum error-correcting codes are typically built on approximative open-quantum-system models such as Born--Markov master equations. However, it is an open question how such codes perform in actual physical systems that, to some extent, necessarily exhibit phenomena beyond the limits of these models. To this end, we employ numerically exact open-quantum-system dynamics to analyze the performance of a five-qubit error correction code where each qubit is coupled to its own bath. We first focus on the performance of a single error correction cycle covering time scales beyond that of Born--Markov models. Namely, we observe distinct power law behavior of the channel infidelity $\propto t^{2a}$: $a\lesssim 2$ in the ultrashort times $t<3/\omega_{\rm c}$ and $a\approx 1/2$ in the short-time range $3/\omega_{\rm c}<t<30/\omega_{\rm c}$, where $\omega_{\rm c}$ is the cutoff angular frequency of the bath. Importantly, the five-qubit quantum-error correction code suppresses all single errors, including those arising from the ultrashort and short-time evolution, which are peculiar to the exact evolution. Interestingly, we demonstrate the breaking points of the five-qubit error correction code and the Born--Markov models for repeated error correction when the repetition rate exceeds $2\pi/\omega$ or the coupling strength $\kappa \gtrsim 0.1 \omega$, where $\omega$ is the angular frequency of the qubit. Our results pave the way for applying numerically exact open-quantum-system models for the studies of QECs beyond simple error models. 
\end{abstract}
\maketitle
\textit{Introduction}.--- One of the significant milestones of quantum computing thus far has been the experimental demonstration of quantum advantage~\cite{arute_quantum_2019,zhong_quantum_2020,wu_strong_2021}, in which a quantum computer solved a logically well-defined computational task faster than usual supercomputers with currently known algorithms. However, the practical use cases are yet to be known since it generally calls for higher qubit numbers and operational fidelity. 
 Whereas it may be possible to improve the number and fidelity of some physical qubits to a level where currently known algorithms~\cite{bharti_noisy_2021,Montanaro_Algorithms} yield practical quantum advantage, the fidelity requirements are thought to be harsh, and hence it seems that completely new algorithms are needed for short-term applications on this path.
 
Quantum error correction (QEC)~\cite{peres_reversible_1985,shor_scheme_1995,knill_theory_1997,la_guardia_quantum_2020} provides an alternative to new algorithms or build-up of extremely high-fidelity qubits by utilizing many physical qubits to build logical qubits of increasingly high fidelity. Notable approaches to qubit-based QEC include the repetition codes~\cite{shor_scheme_1995}, surface codes~\cite{ fowler_high-threshold_2009,fowler_surface_2012, Gottesman1997StabilizerCA}, and colour codes~\cite{fowler_two-dimensional_2011}. An alternative is to encode quantum information in bosonic modes through the bosonic codes~\cite{Chuang_bosoniccodes_PRA_1997,Michael2016_bosoniccodes}.
Thus the path of QEC seems theoretically more established~\cite{knill_scheme_2001,fowler_high-threshold_2009,fowler_two-dimensional_2011,fowler_surface_2012,Laflamme1996} than that of noisy intermediate scale quantum advantage, which has motivated a number of milestone experiments ranging from the early~\cite{cory_experimental_1998,chiaverini_realization_2004} and more recent~\cite{schindler_experimental_2011,cramer_repeated_2016,riste_detecting_2015,kelly_state_2015,chen_exponential_2021} demonstrations of either bit or phase flip codes to single~\cite{abobeih_fault-tolerant_2022,egan_fault-tolerant_2021} and repeated error detection~\cite{andersen_repeated_2020,chen_exponential_2021,marques_logical-qubit_2022} and correction~\cite{ryan-anderson_realization_2021,krinner_realizing_2022,zhao_realization_2022,sundaresan_matching_2022,acharya_suppressing_2022}. 

Most of the experimental works have focused on single logical qubits, but very recently also an entangling gate between two logical qubits was demonstrated~\cite{Ryan-Anderson_EntanglingGates_2022}. Nevertheless, the experimental tests of the scaling of logical errors with increasing size of the code size is still at its infancy, and importantly, it remains to be verified whether the relatively simple error models used in most theoretical studies of QEC are adequate. Namely, the usual local Born--Markov approach to open quantum systems~\cite{Heinz-PeterBreuer2007} motivates an error model where Poisson-distributed bit and phase flips are applied on the individual physical qubits~\cite{Lidar2014}. However, non-Markovian dynamics~\cite{breuer_colloquium_2016}, global effects of the environment on the system~\cite{VasiliiPRB,Teixeira2021}, and system--environment correlations~\cite{tuorila_system-environment_2019,PhysRevX_sahar} are neglected in such approaches, which raises the concern whether these phenomena can lead to small but significant error in the logical qubits that need to operate at extremely high fidelity. 

To analyze the performance of error correction codes beyond the simple models, we employ a numerically exact treatment of the open-quantum-system dynamics arising from five physical qubits, each coupled to their Ohmic baths with the second-order Lorentz--Drude cut-off at angular frequency~$\omega_\textrm{c}$. 
We encode the qubits into a five-qubit QEC code~\cite{Laflamme1996} and first assess the channel induced by the numerically exact model for a single error correction cycle.   At ultrashort times $t< 3/\omega_\textrm{c}$ and short times $3/\omega_\textrm{c}<t<30/\omega_\textrm{c}$, we observe apparent differences in the channel fidelity corresponding to the exact and Born--Markov dynamics. We further validate our conclusions with an analytic model for the short-time dynamics. We also apply repeated error correction and observe deviations from the infidelity given by  Born--Markov results.  The  Born--Markov results agree with other methods when the repetition rate does not exceed $2\pi/\omega$ or the coupling strength  $\kappa \gtrsim 0.1 \omega$.  Thus, in experimentally relevant scenarios of Rabi-driven qubits, the Born--Markov approach, and hence the usually employed error model of bit and phase flips, seem feasible to describe the quantum memory protected by the five-qubit code.\\

\begin{figure}[t]
    \centering
    \includegraphics[scale=1.0]{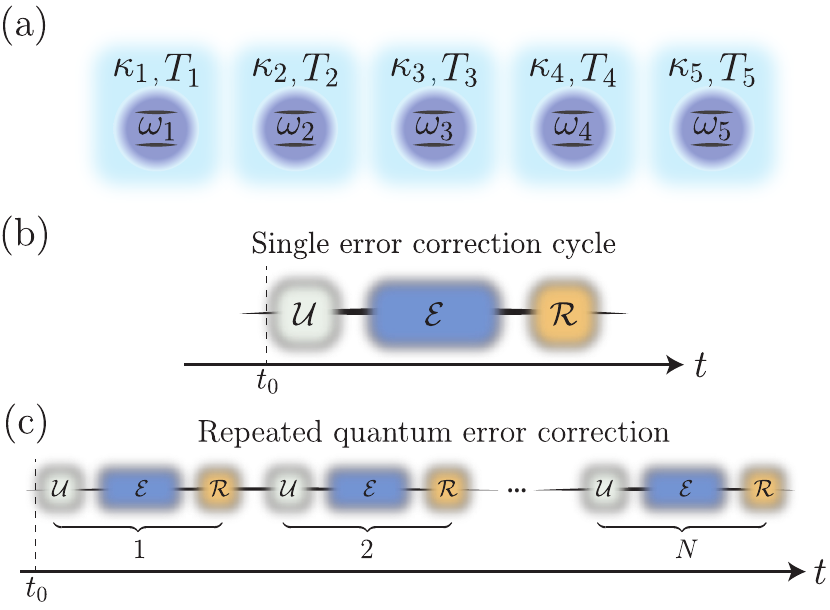}
    \caption{(a) Schematic of the five-qubit setup, where each qubit is coupled to its own Ohmic bath. Here $T_j$ is the temperature of the bath, $\kappa_j$ defines the strength of interaction, and $\omega_j$ is the qubit frequency. (b) Schematic of the single-cycle error correction process where $\mathcal{U}$ is the encoding process, $\mathcal{E}$ is the error process and $\mathcal{R}$ is the recovery operation. (c) Schematic of the repeated error correction process. Note that here the recovery operation $\mathcal{R}$ also includes the decoding process.}
    \label{fig1}
\end{figure}

\textit{Open many-body quantum system }.---We consider a system of five non-interacting qubits, each of them is coupled to its own decay channel modelled as a bath of bosonic modes shown in Fig.~\ref{fig1}(a). The total system-bath Hamiltonian reads as $\hat H=\sum_{j=1}^5 \hat H_{j}$ with
\begin{align}
    \frac{\hat H^{}_j}{\hbar}  = -\frac{\omega_j}{2} \hat \sigma^{}_{\mathrm z j} + \sum_{k}\left[ \Omega^{}_{kj} \hat n^{}_{kj}+ g^{}_{kj} \left(\hat b^{}_{kj} + \hat b_{kj}^\dag\right) \hat \sigma^{}_{\mathrm x j}  \right], 
    \label{eq:Hamiltonian}
\end{align}
where $\omega_j$ and $\hat{\sigma}_{\mathrm \alpha j}$ ($\mathrm \alpha=\mathrm x,\mathrm y,\mathrm z$) are the angular frequency and Pauli matrices of the qubit $j$,  respectively, and  $\Omega_{kj}$ is the angular frequency of the mode $k$ in the $j$-th bath, $\hat{b}_{kj}$ and $\hat{n}_{kj}=\hat{b}^{\dagger}_{kj}\hat{b}^{}_{kj}$ are its annihilation and number operators, and $g_{kj}$ is its coupling strength of the qubit $j$. We assume that the errors are fully described by the interaction between the qubits and their individual baths. 
The effect of each local bath on the corresponding qubit can be completely described by its temperature $T_j$ and the spectral density~\cite{Weiss2008}
\begin{equation}
    J_j(\Omega) = \pi \sum\limits_k g^2_{kj} \delta(\Omega - \Omega^{}_{kj}),
\end{equation}
where $\delta(\Omega-\Omega^{}_{kj})$ is the Dirac's delta function peaked at $\Omega^{}_{kj}$. We assume an ohmic-type distribution 
$J_j(\Omega) = (\kappa_j /\omega_j)\Omega/\big(1+\Omega^2/\omega^2_{\mathrm c j}\big)^2$, where $\kappa_j$ is the coupling strength of the qubit $j$ to its bath and $\omega_{\mathrm c j}$ is the bath cutoff frequency.

An open many-body system  dynamics can be quite difficult to analyse in general. Typically the Lindblad master equation method ~\cite{Heinz-PeterBreuer2007, SM} is utilised to simulate the system dynamics.   
This approach is popular for its simplicity but it is justified only for weak coupling between the qubits and their baths, short bath correlation time which restricts the bath temperature from below, and high energy separation of the qubit states as compared to the resulting level broadening.

For accurate simulations at strong coupling strengths or at short time scales, one needs to employ numerically exact methods ~\cite{Stockburger1999,tuorila_system-environment_2019,VasiliiPRB, Babu2021}. Here, we use the stochastic Liouville equation with dissipation (SLED) which for our multiqubit system~\eqref{eq:Hamiltonian} takes the form~\cite{stockburger_stochastic_1999,tuorila_system-environment_2019}
\begin{multline}
    \frac{\mathrm d\hat \rho}{\mathrm dt} = \sum\limits_{j=1}^5\left\{
    i \frac{\omega_{j}}{2} \left[\hat \sigma_{\mathrm z j}, \hat \rho\right] + i \kappa_{j}\left[\hat \sigma_{\mathrm x j}, \left\{\hat \sigma_{\mathrm y j},  \hat \rho\right\}\right] \right.
    \\ \left .-\frac{\kappa_j}{\hbar \omega_j\beta_j}
    \left[\hat\sigma_{\mathrm x j} \left[\hat \sigma_{\mathrm x j}, \hat \rho\right]\right]
    - i\xi_j(t) \left[\hat \sigma_{\mathrm x j},\hat \rho\right]
    \right\},
    \label{sled}
\end{multline}
where $\beta_j=1/(k_{\rm B}T_j)$, and the colored real-valued Gaussian noise $\xi_j(t)$ has the correlation function
\begin{multline}
    \left\langle \xi_j(t) \xi_j(0)\right \rangle
    = \\ \frac{1}{\pi}\int\limits_0^{+\infty} J_j(\Omega)
    \left[\coth\left(\frac{\hbar \Omega\beta_j}{2}\right) - \frac{2}{\hbar \Omega\beta_j}\right]\cos \left(\Omega t\right)\;\mathrm d\Omega .
    \label{eq:noise}
\end{multline}
It is important that the full spectral density with the proper cut-off should be considered in  Eq.~\eqref{eq:noise}, not just its low-frequency Ohmic asymptotics. Note that SLED is a stochastic equation, and thus the final result is an average over the solutions of Eq.~\eqref{sled} for several noise realizations.

\begin{figure*}[t]
    \includegraphics[width=0.49\linewidth]{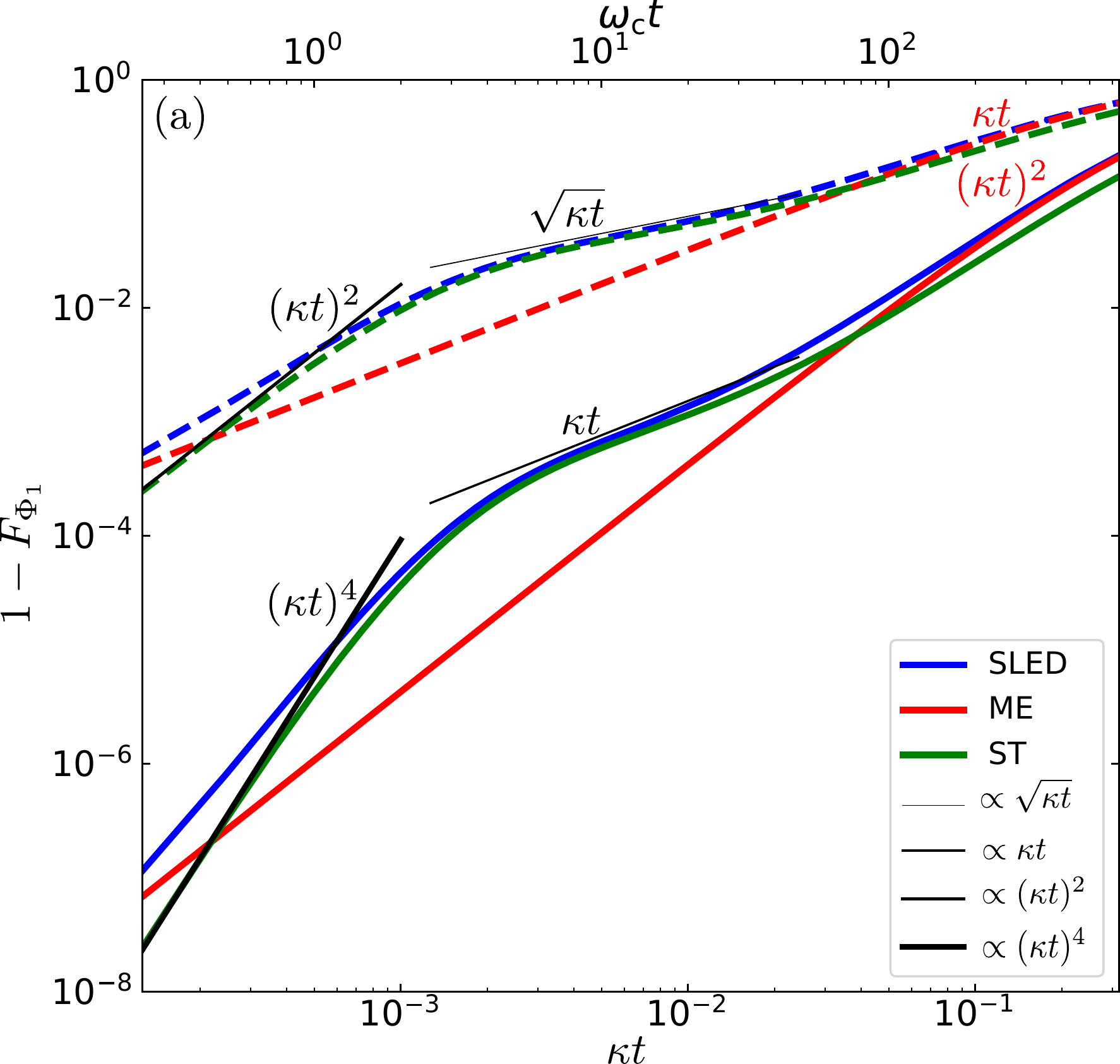}
    \hspace{0.1 cm}
    \includegraphics[width=0.49\linewidth]{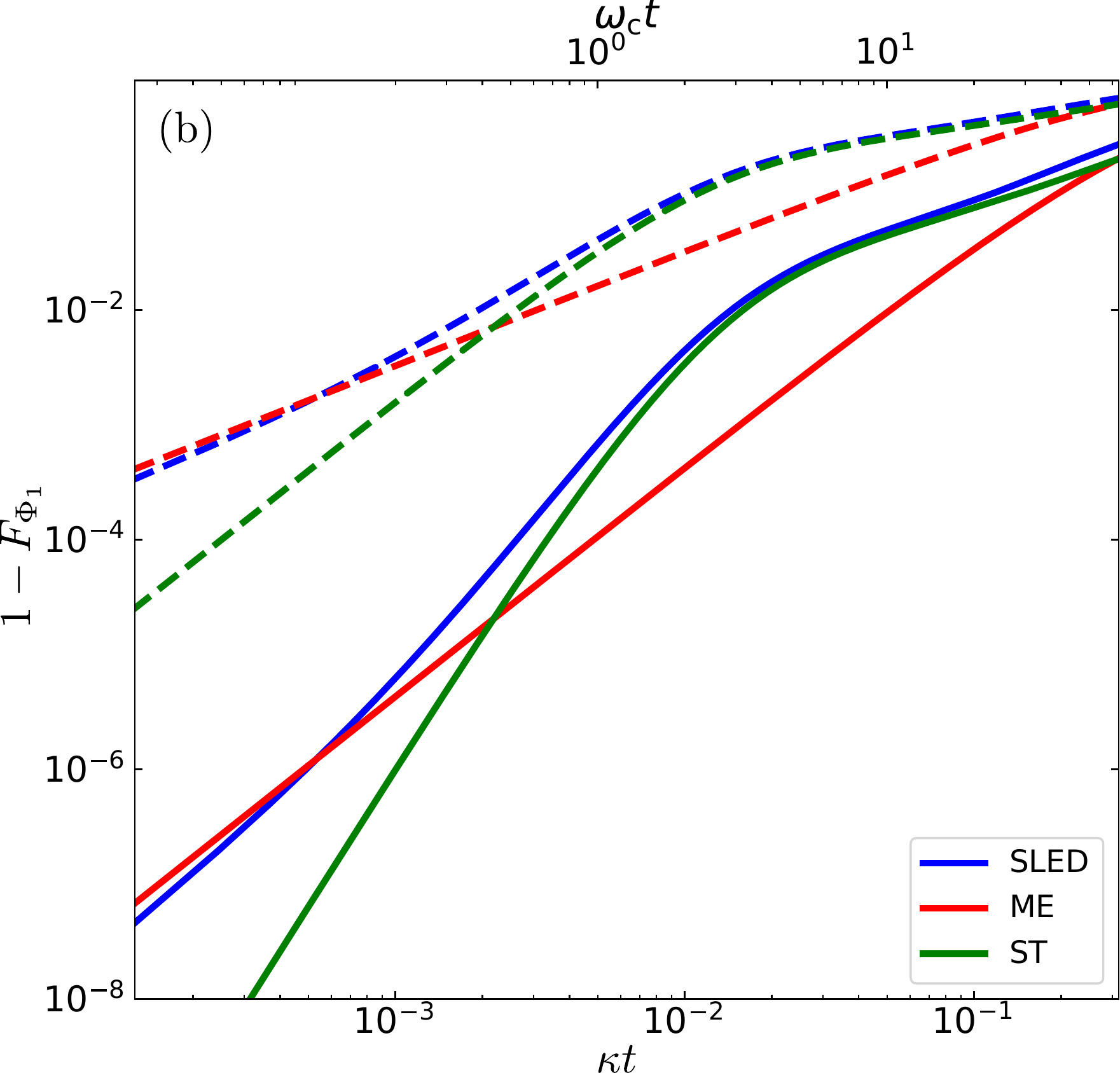}
     \caption{Channel infidelity $1-F_{\Phi_1}$ for the non-error corrected channel (dashed lines) and the error-corrected channel (solid lines) as a function of the time interval $\kappa t$ from the encoding to the recovery, i.e. the duration of the error process for coupling strength (a)~$\kappa/\omega=0.01$ and (b)~$\kappa/\omega=0.1$. We compare here results calculated by SLED (blue), the Lindblad master equation (ME, red), and the analytic short-time methods by Eq.~\eqref{short-time eq.}  (ST, green). The qubits and baths are assumed identical with $\omega_{j }=\omega$, $\kappa_j=\kappa$, and $\beta_j=\beta=2$ and  Drude-type cut-off at the frequency $\omega_{\rm c j}/\omega=20$. 
     }
    \label{fig2}
\end{figure*}

\textit{Quantum error correction}.---To correct the errors on each qubit caused by their own local baths, we employ the five-qubit error correction code introduced
in Ref.~\cite{Laflamme1996}. We choose this code since it uses the minimum number of physical qubits to perfectly correct arbitrary single-qubit errors. The code operates with choosing a single main qubit such as $j=3$ and adding four other qubits, e.g. $1,$ $2$,
$4$ and $5$. A single error correction cycle is characterized by an
encoding process $\mathbf{\mathcal{U}}$ at the beginning of the time-evolution, the error dynamics $\mathcal{E}$ followed by the recovery operation $\mathcal{R}$ as depicted in Fig. 1(b). Here,
$\mathcal{U}$, $\mathcal{E}$, and $\mathcal{R}$ are superoperators
acting on the five-qubit system. The corresponding logical states and the Kraus map associated with $\mathcal{R}$ are shown in Ref.~\cite{SM}.

The recovery $\mathcal{R}$ comprises the error detection through
$\mathcal{U}^{-1}$ and local measurements of the ancillae, their  outcomes of which determine a unitary operation on the main qubit~\cite{Laflamme1996}. Notice that the chosen recovery process uses no extra ancillae beyond the qubits $1$, $2$, $4$, and $5$. 
In realistic implementations, repeated cycles of
error correction are usually required, see Fig. 1(c).

The five-qubit dynamics
in the presence of repeated error correction can thus be ascribed
to the quantum map 
\begin{equation}
\Phi_{\textsc{N}}=\left(\mathcal{R}\mathcal{E}\mathcal{U}\right)^{N},\label{eq:Phi}
\end{equation}
where $N$ is the number of error correction cycles. We emphasise
that our focus here is to probe the deleterious effects on quantum
error correction in the presence of an accurate description of error dynamics $\mathcal{E}$ promoting versatile decoherence, typically neglected in the error models, and not to model actual physical qubit possibly affected by several additional error sources typically encountered in NISQ devices such as  gate and SPAM errors~\cite{Georgopoulos2021}. Thus, we assume instantaneous $\mathcal{U}$ and $\mathcal{R}$
hereafter.

The performance of the five-qubit QEC is analysed through the channel fidelity~\cite{Schumacher1996,Albert2018} 
\begin{equation}
F_{\Phi_{\textsc{N}}}=\frac{1}{8}\sum_{j=0}^{3}\Tr\left[\hS_{j}\Phi_{\textsc{N}}\left(\hS_{j}\right)\right],\label{eq:chfid}
\end{equation}
where $\hS_{0}  =\kb{0_{L}}{0_{L}}+\kb{1_{L}}{1_{L}}$, $\hS_{1}  =\kb{0_{L}}{1_{L}}+\kb{1_{L}}{0_{L}}$, $\hS_{2}  =-i\kb{0_{L}}{1_{L}}+i\kb{1_{L}}{0_{L}}$, and $\hS_{3}  =\kb{0_{L}}{0_{L}}-\kb{1_{L}}{1_{L}}$ are the identity and Pauli operators in the logical qubit subspace $\left\{ \ket{0_{L}},\ket{1_{L}}\right\}$. The channel fidelity effectively quantifies the success of quantum information preservation under the action of $\Phi_{\textsc{N}}$, so that successful QEC should produce values close to unity.\\

\textit{Single cycle quantum error correction}.---We first analyze a single error correction cycle of the five-qubit code and compute the channel fidelity with the SLED and the Lindblad models. For simplicity, we assume that all the qubits and the corresponding baths are identical with $\omega_{j }=\omega$, $\kappa_j=\kappa$, and $\beta_j=\beta$, for $j \in \{1, \dots 5\}$. Then, in Fig.~\ref{fig2}, we show the infidelity $1-F_{\Phi_1}$ of  the quantum channel in Eq.~\eqref{eq:Phi} without the recovery operation $\mathcal{R}$, i.e. the error channel infidelity (dotted lines),  and the full channel that includes the recovery operation (solid lines) for  $ \kappa/\omega  =0.1$ (a), and  $ \kappa/\omega =0.01$ (b).

The error channel infidelity calculated with the Lindblad model shows a linear dependence on $\kappa t$ as the single-qubit error probability per error correction cycle is proportional to $\kappa t$. Ideally, the five-qubit error correction protocol corrects all the first-order single-qubit errors, i.e. a single Pauli operator action on single-qubit, and leaves the higher-order errors uncorrected. Thus, the infidelity of the recovered channel essentially includes those uncorrected errors, and the recovered channel infidelity calculated with the Lindblad model is proportional to $(\kappa t)^2$.

The numerically exact dynamics calculated with SLED considerably differ from the Lindblad predictions as shown in Fig.~\ref{fig2}(a). 
In ultrashort times $\omega_{\rm c}t \le 3$, the error channel infidelity is proportional to $(\kappa t)^{a}$, where $a  \le 2$. Whereas  in short times, $3<\omega_{\rm c}t <30$, the error channel infidelity is proportional to $(\kappa t)^{b}$, where $b \approx 1/2$. For the long time limit, $\omega_{\rm c} t>10$, the error channel infidelity shows a linear dependence on $\kappa t $ as same as the Lindblad results. Again the recovery process corrects the first-order errors and leaves the second-order errors uncorrected. Thus, recovery channel infidelity is proportional to:  $(\kappa t)^{2a}$ at ultrashort times, $(\kappa t)^{2b}$ at short times and $(\kappa t)^{2}$ at long times. Fig. \ref{fig2}(b) shows infidelity estimates for relatively strong coupling, i.e. $\kappa/\omega =0.1 $. Similar to the previous case, SLED exhibits deviations in ultrashort and short times. However, the channel infidelity arising from this short-time dynamics at $t\approx 1/\omega_\textrm{c}$ appears to be ten times greater than that of the case in Fig.~\ref{fig2}(a) with $\kappa/\omega =0.01$.

Deviation of the SLED results at short times is due to the universal decoherence, where the intrinsic dynamics of the system stays essentially frozen  and the high-frequency reservoir modes control the system dynamics~\cite{braun_universality_2001,tuorila_system-environment_2019,Babu2021}. 
We can obtain the resulting Liouvillian superoperator $\mathcal{L}^{\text{ST}}(\hat \rho)=\mathrm d \hat \rho/\mathrm dt$ of dynamics as 
\begin{equation}
    \mathcal{L}^{\text{ST}}(\hat \rho)=\sum_{j=1}^5\frac{f_j'(t)\kappa_j }{\omega_j\pi }(\hat \sigma_{\mathrm xj}\hat \rho \hat \sigma_{\mathrm xj}- \hat \rho),
    \label{short-time eq.}
\end{equation}
where $f_j'(t)$ is the time derivative of the integral average function $f_j(t)$, defined as
\begin{align}
    f_j(t)=\frac{2\omega_j}{\kappa_j}\int_0^{\infty}\mathrm d\Omega \frac{J_j(\Omega)}{\Omega^2} \coth(\hbar\beta_j\Omega/2)\cos(\Omega t/2).
\end{align}
Here, we utilise an extended version of the single qubit short-time dynamics presented in Refs.~\cite{tuorila_system-environment_2019, Babu2021} for the derivation of the time evolution operator,  see Ref.~\cite{SM} for more details. 

In ultrashort times, $t<3/\omega_{\rm c}$, $f(t) \approx -\omega_{\rm c}^2 t^2/2 $ and  we can obtain the fidelity of error channel as $F_{\Phi_1} = [1+\exp(-\kappa \omega_{\rm c}^2t^2/\omega \pi)]^5/32$. Thus, infidelity $1-F_{\Phi_1} $ is proportional to $(kt)^2$ in ultrashort times. At later times, $t\leq 30/( \omega_{\rm c})$ infidelity shows roughly $\sqrt{\kappa t }$ behavior.  The infidelity estimates with the analytic model are represented with green lines in Fig.~\ref{fig2}. 
The SLED results closely follow the analytic dynamics, although there are some deviations in ultrashort times. 

Our results demonstrate that the SLED-based simulation seems adequate in studying the performance of quantum error correction codes beyond the typical simple error models. We emphasise that the short-time and the long-time error can be corrected to a large extent with a single recovery cycle of the five-qubit code: the dominant first-order errors are corrected, rendering the second order processes to dominate the remaining error channel subsequently. However, in the short times $3<\omega_{\rm c} t <30$, the errors occur so frequently that recovery process of the distance d=3 error correction code is not suppressing them enough, visualized best in the repeated error correction protocol.\\  

\begin{figure}[ht]
    \centering
    \includegraphics[width=1\linewidth]{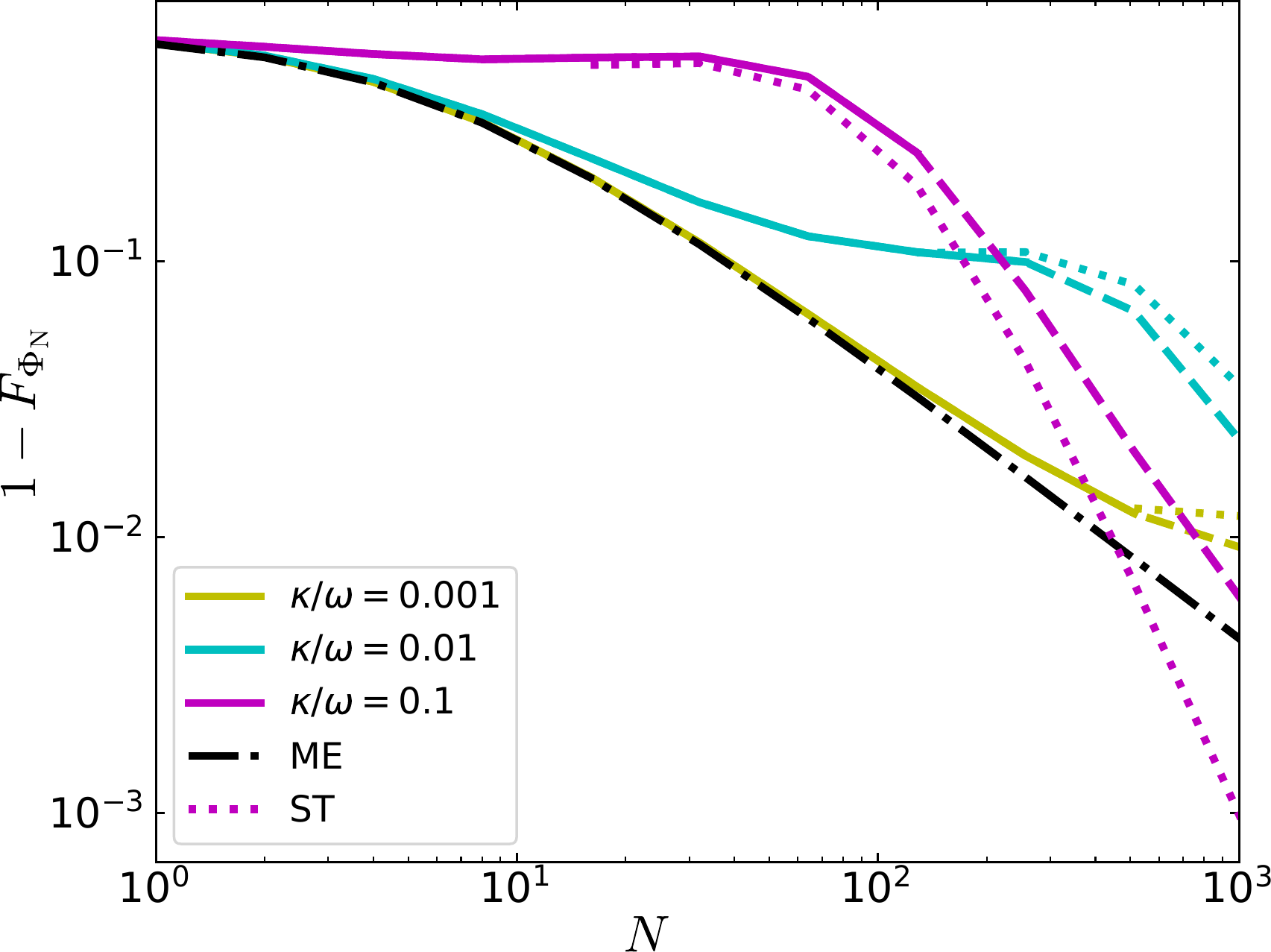}
     \caption{Channel infidelity of the repeated error correction $1-F_{\Phi_{\textsc{N}}}$ as a function of the number of the error correction cycles $N$ within a fixed total time $t_{\rm max}\kappa=1$ for different coupling strengths $\kappa/\omega=0.001$ (yellow), $0.01$ (blue), and $0.1$ (magenta). The solid and dashed lines are computed by the SLED and the dotted lines show the results by the analytic short-time time-evolution by Eq.~\eqref{short-time eq.}. The black dash-dotted line represents the infidelity computed by using the Lindblad master equation. The Lindblad solutions are the identical for all  values of $\kappa$. }
    \label{fig3}
\end{figure}

\textit{Repeated quantum error correction}.---Finally, we study what happens to the five-qubit quantum error correction at strong coupling $\kappa$, and find how strong environmental coupling is enough to demolish the benefits of active error correction. To this end, our focus is on the scheme of repeated quantum error correction, visualized in Fig.~\ref{fig1}(c), where we fix the total time interval $t_{\rm max}=\kappa^{-1}$ and vary the number of error correction cycles $N$ within this interval. The expectation is that for a well-functioning error-correction process, an increase in the number of cycles $N$ decreases the infidelity for the final state at~$t_{\rm max}$.\\

Considering the Lindbladian description of the error process, 
the probability of the dominant uncorrected error per cycle is proportional to $(\kappa t)^2$, yielding that the channel infidelity at $\kappa t_{\rm max}=1$ scales as $N^{-1}$ for $N\gg1$ independent on the value of $\kappa$ as shown in Fig.~\ref{fig3}.  To explore the effect of strong coupling, we compute in Fig.~\ref{fig3} the final infidelity of the repeated error correction process using SLED at the limit of weak, moderate  and strong coupling strengths, corresponding to $\kappa/\omega=0.001$, $0.01$, and $0.1$. We observe that at the weak coupling limit, the SLED infidelity shows only minor deviations from the scaling behaviour predicted with the Lindblad model. Unsurprisingly, quantum-error correction by the perfect five-qubit code functions adequately at weak qubit-environment coupling $\kappa/\omega\lesssim 0.001$. However, already at moderate values $\kappa/\omega=0.01$, the infidelity starts to show plateauing as a function of $N$ for $10\lesssim  N \lesssim 100$. The situation is even worse by strong coupling $\kappa/\omega=0.1$ showing complete plateauing for $N\lesssim 100$. In both cases, the plateauing arises when the time interval between the recovery operations $\delta t =t_{\rm max}/ N$ is in the short-time domain where the error probability scales $(\kappa t)^{1/2}$. This result indicates that a simple distance-three error correction cannot overcome the fast occurrence of errors with probability~$(\kappa t)^{1/2}$.

When the repetition rate of error correction is very fast, corresponding to the ultrashort-time dynamics, we observe very favorable scaling of the infidelities as a function of $N$. However, the ultrashort dynamics corresponds the case where the repetition rate of error correction is much faster than the qubit frequency, which contradict the quantum speed limit . For example, for $\kappa/\omega=0.1$ the favorable regime starts when $N > 10^2$ which yields $t<t_{\rm max}/N=\omega^{-1}/10$.\\

\textit{Conclusions \& Discussion}.---Typical quantum error correction methods assume error models based on Born-Markov assumptions. In this work, we have subjected these assumptions to detailed scrutiny seeking the fundamental limits of the quantum error correction processes. To this end, we have analyzed the performance of the five-qubit error correction code using a numerically exact open-quantum-system error model. 
We observed variations in the resulting infidelity from the typical power law behavior predicted by the Born--Markov model, specifically in the short-time domain $3<\omega_{c}t\leq 30$ and ultrashort times where $\omega_{c}t\leq 3$. These deviations arise from short-time universal decoherence induced by the bath modes. 
All the single-qubit errors arising from the interaction with environment are still correctable. However, at short times, the errors occur so frequently that the five-qubit error correction code becomes ineffective. 
We also substantiated the SLED result with an analytical error model incorporating short-time universal decoherence. Furthermore, we analyzed the repeated error correction process and observed fidelity improvement in some operational regimes. However, the five-qubit error correction code fails to yield improved fidelities when the repetition rate exceeds $2\pi /\omega$, where $\omega$ is the angular frequency of the qubit. Fast recovery operations with a repetition rate greater than  $100 \times 2\pi /\omega$  can result in very low infidelity, but this regime does not seem physically feasible. Finally, we demonstrated the breaking of the Born-Markov method when the repetition rate exceeds $2\pi /\omega$ or the coupling strength $\kappa \gtrsim 0.1 \omega$.

The consequences of the universal decoherence occur at such short time scales that they are beyond the current experimental state-of-the-art and yield practically no no-go results for typical error correction codes. An interesting future research topic is the combined effect of qubit-qubit crosstalk and universal decoherence from the point of view of quantum error correction. Our results demonstrates the feasibility of employing numerically exact open-quantum-systems methods to analyze the performance of the QECs. Further studies beyond simple error models may benefit the future development of QEC codes. The numerically exact method we employed here may also find applications in simulating other many-body systems strongly coupled to the environment. \\

\textit{Acknowledgments}.---We  thank Jani Tuorila, Tapio Ala-Nissila, J\"{u}rgen Stockburger, and Joachim Ankerhold for useful discussions. We acknowledge funding by Scientific Advisory Board for Defence (MATINE), Ministry of Defence of Finland, European Research Council under Consolidator Grant No. 681311 (QUESS) and Advanced Grant No. 101053801 (ConceptQ), and the Academy of Finland under Grants Nos. 316619 and 336810. The  authors wish to acknowledge CSC -- IT Center for Science, Finland, for computational resources.
%

%
\onecolumngrid
\newpage
\widetext
\section{Supplemental Material} 
\subsection{Details of Encoding and Recovery protocol for the five--qubit code}

In five-qubit code, the encoding $\mathcal{U}$ uses
a sequence of nonlocal gates to encode the state of the main qubit
into the logical subspace $\left\{ \ket{0_{L}},\ket{1_{L}}\right\} $,
where
\begin{align}
\ket{0_{L}} =& \left(\ket{0}+\ket{6}+\ket{9}-\ket{15}-\ket{19}+\ket{21}+\ket{26}+\ket{28}\right)/\sqrt{8},\nonumber\\
\ket{1_{L}} = &\left(-\ket{3}-\ket{5}-\ket{10}+\ket{12}-\ket{16}+\ket{22}+\ket{25}+\ket{31}\right)/\sqrt{8},\label{eq:cdwd}
\end{align}
and $\left\{ \ket{i},\ i=0,1,\dots,31\right\} $ is the five-qubit computational
basis. Here, we have used the decimal representation of the binary numbers $i=(i_1,i_2,i_3,i_4, i_5)_2$, $i_j\in\{0,1\}$, associated to the tensor product states $\{\ket{i_1\dots i_5}\}$. 
A single recovery stage considered in the simulations of the main text can be written
as the following Kraus map 
\begin{equation}
\mathcal{R}(\hrh)=\sum_{k=0}^{15}\hR_{k}\hU^{\dagger}\hrh\hU\hdgg R_{k},\label{eq:recKraus}
\end{equation}
where $\hU$ is the unitary operator associated to the encoding process
$\mathcal{U}$ yielding the codewords $\ket{0_{L}}$ and $\ket{1_{L}}$,
see Ref.~\cite{Laflamme1996}. In Eq.~\eqref{eq:recKraus}, the operators $\left\{ \hR_{k}\right\} $
account for projective measurements of the auxiliary qubits $1$, $2$,
$4$, and $5$ in the computational basis, whose outcomes drive a unitary operation to correct the state of the main qubit $3$. To carry on the next recovery cycles, the auxiliary qubits are then reset to the state $\ket{0000}_{1245}$. Therefore, the operators $\left\{ \hR_{k}\right\}$ can be
explicitly written as
\begin{align}
\hR_{0} & =\ket{00}_{12}\bra{00}\otimes\hI_{3}\otimes\ket{00}_{45}\bra{00},\,\,
\hR_{1}  =\ket{00}_{12}\bra{00}\otimes\hsg_{\mathrm z3}\otimes\ket{00}_{45}\bra{01},\,\,
\hR_{2}  =\ket{00}_{12}\bra{00}\otimes\hI_{3}\otimes\ket{00}_{45}\bra{10},\nonumber\\
\hR_{3} & =\ket{00}_{12}\bra{00}\otimes\hI_{3}\otimes\ket{00}_{45}\bra{11},\,\,
\hR_{4}  =\ket{00}_{12}\bra{01}\otimes\hI_{3}\otimes\ket{00}_{45}\bra{00},\,\,
\hR_{5}  =\ket{00}_{12}\bra{01}\otimes\hsg_{\mathrm z3}\otimes\ket{00}_{45}\bra{01},\nonumber\\
\hR_{6} & =\ket{00}_{12}\bra{01}\otimes\hsg_{\mathrm x3}\otimes\ket{00}_{45}\bra{10},\,\,
\hR_{7}  =\ket{00}_{12}\bra{01}\otimes\hsg_{\mathrm x3}\otimes\ket{00}_{45}\bra{11},\,\,
\hR_{8}  =\ket{00}_{12}\bra{10}\otimes\hI_{3}\otimes\ket{00}_{45}\bra{00},\nonumber\\
\hR_{9} & =\ket{00}_{12}\bra{10}\otimes\hsg_{\mathrm x3}\otimes\ket{00}_{45}\bra{01},\,\,
\hR_{10}  =\ket{00}_{12}\bra{10}\otimes\hsg_{\mathrm z3}\otimes\ket{00}_{45}\bra{10},\,\,
\hR_{11}  =\ket{00}_{12}\bra{10}\otimes\hsg_{\mathrm x3}\otimes\ket{00}_{45}\bra{11},\nonumber\\
\hR_{12} & =\ket{00}_{12}\bra{11}\otimes\hsg_{\mathrm z3}\otimes\ket{00}_{45}\bra{00},\,\,
\hR_{13}  =\ket{00}_{12}\bra{11}\otimes\hsg_{\mathrm x3}\hsg_{\mathrm z3}\otimes\ket{00}_{45}\bra{01},\,\,
\hR_{14}  =\ket{00}_{12}\bra{11}\otimes\hsg_{\mathrm x3}\otimes\ket{00}_{45}\bra{10},\nonumber\\
\hR_{15} & =\ket{00}_{12}\bra{11}\otimes\hsg_{\mathrm z3}\otimes\ket{00}_{45}\bra{11},\label{eq:recKraus2}
\end{align}
where $\hI_{3}$ is the identity operator and $\hsg_{\mathrm\alpha 3}$ ($\mathrm\alpha=\mathrm x,\mathrm y,\mathrm z$) are the Pauli matrices for qubit $3$.
\subsection{Lindblad master equation for the Five-qubit system}
Commonly used Born--Markov--secular approximation reduces the joint unitary evolution of extended system of qubits and their baths to the non-unitary evolution of the reduced density operator $\hat \rho$ of the five-qubit system governed by the Lindblad master equation~\cite{Heinz-PeterBreuer2007}
\begin{align}
    \frac{\mathrm d \hat \rho}{\mathrm dt} = \sum\limits_{j=1}^5 \{
    i \frac{\omega_{j}}{2} [\hat \sigma_{\mathrm z j}, \hat \rho] + 
    \kappa_j [n_j\left(\omega_{j}) + 1\right] \mathcal D\left[\hat \sigma_{-j}\right] \hat \rho +
    \kappa_j n_j(\omega_{j}) \mathcal D\left[\hat \sigma_{+j}\right] \hat \rho
    \},
\end{align}
where $\mathcal D[\hat A] \hat \rho = 2 \hat A \hat \rho \hat A^\dag - \hat A^\dag \hat A \hat \rho - \hat \rho \hat A^\dag \hat A $, and $n_j(\Omega) = \left[1 - \exp\left(-\beta_j\hbar \Omega \right)\right]^{-1}$ is the average thermal occupation number of the $j$-th bath with $\beta_j=1/(k_{\rm B}T_j)$. 
\subsection{Analytic model for the short-time error dynamics}
The short-time decoherence in the case of a single qubit interacting with a bosonic bath has already been demonstrated in Ref.s \cite{tuorila_system-environment_2019,Babu2021}. Let us first analyse a single qubit in the five-qubit setting and extend it to the whole system later. In the early time limit, system dynamics remain frozen, and high-frequency environmental modes control the dynamics. Thus, we completely ignore the system Hamiltonian  and  obtain the elements of the reduced density matrix in the eigenbasis of the operator $\hat \sigma_{\mathrm x j} $ as  \cite{tuorila_system-environment_2019,Babu2021}
\begin{align}
  \langle n|\hat\rho^{j}(t)|m\rangle= \langle n|\hat\rho_{\rm S}(0)|m\rangle \exp( &[-(n-m)^2f_j(t)+i(n^2-m^2)\phi_j(t)]\kappa_j/2 \pi \omega_j ) .
\end{align}
where $f_j(t)$ and $\phi_j(t)$ are integral average functions and  which takes the forms of
\begin{align}
   & f_j(t)=\frac{\omega_j}{\kappa_j}\int_0^{\infty}d\Omega \frac{J_j(\Omega)}{\Omega^2} \coth(\hbar\beta_j\Omega/2)[1-\cos(\Omega t)],\nonumber\\&
   \phi_j(t)=\frac{\omega_j}{\kappa_j}\int_0^{\infty}d\Omega \frac{J_j(\Omega)}{\Omega^2} [\Omega t-\sin(\Omega t)].
\end{align}
Using the expansion of $\hat \sigma_{\mathrm xj} =\sum_n n \ket{n}\bra{n}$, we can write the early time evolution operator $\mathcal{L}^{\text{ST}}(\hat \rho^j)$ as
\begin{align}
    \mathcal{L}^{\text{ST}}(\hat \rho^j)=\frac{i \phi'(t)\kappa }{2\pi \omega_j}[(\hat \sigma_{\mathrm x j})^2,\hat \rho ] +\frac{f_j'(t)\kappa_j }{2\pi\omega_j }[2\hat \sigma_{\mathrm xj} \hat \rho^j \hat \sigma_{\mathrm xj}-(\hat \sigma_{\mathrm xj})^{2} \hat \rho^j- \hat \rho^j (\hat \sigma_{\mathrm xj})^{2}]
\end{align}
where $\mathcal{L}^{\text{ST}}(\hat \rho^j)=\mathrm d\hat \rho^j/\mathrm d t$. We can then write the time evolution operator for the whole system as
\begin{align}
    \mathcal{L}^{\text{ST}}(\hat \rho)=\sum _{j=1}^5 \frac{f_j'(t)\kappa_j }{\pi\omega_j }\big(\hat \sigma_{\mathrm xj} \hat \rho \hat \sigma_{\mathrm xj}-\hat \rho \big )
\end{align}
\subsection{Numerical time integration for SLED}
We use stochastic Liouville equation with dissipation (SLED) to simulate numerically exact error dynamics,  which takes the form~~\cite{stockburger_stochastic_1999,tuorila_system-environment_2019}
\begin{align}
    \frac{\mathrm d\hat \rho}{\mathrm dt} = \sum\limits_{j=1}^5\left\{
    i \frac{\omega_{j}}{2} \left[\hat \sigma_{\mathrm z j}, \hat \rho\right] + i \kappa_{j}\left[\hat \sigma_{\mathrm x j}, \left\{\hat \sigma_{\mathrm y j},  \hat \rho\right\}\right] -\frac{\kappa_j}{\hbar \omega_j\beta_j}
    \left[\hat\sigma_{\mathrm x j} \left[\hat \sigma_{\mathrm x j}, \hat \rho\right]\right]
    - i\xi_j(t) \left[\hat \sigma_{\mathrm x j},\hat \rho\right]
    \right\},
\end{align}
where the colored real-valued Gaussian noise $\xi_j(t)$ has the correlation function
\begin{align}
    \langle \xi_j(t) \xi_j(0) \rangle
    =  \frac{1}{\pi}\int\limits_0^{+\infty} J_j(\Omega)
    \Big[\coth\big(\frac{\hbar \Omega\beta_j}{2}\big) - \frac{2}{\hbar \Omega\beta_j}\Big]\cos \Omega t)\;\mathrm d\Omega . 
    \label{eq:noise1}
\end{align}
The problem can be cast into the form
\begin{equation}
    \frac{d}{dt}\vec v(t) = \mat M(t)\vec v(t),
\end{equation}
where~$\vec v(t)$ is the unknown vector, and~$\mat M(t)$ is the matrix determining the problem. In our case~$\vec v(t)$ is the vectorized form of the density operator, and~$\mat M(t)$ is the matrix form of the superoperator defining the SLED. If one knows the vector~$\vec v(t)$ at time~$t$, the solution after a short time step~$\delta t$ can be obtained with the Magnus expansion,
\begin{equation}
    \vec v(t+\delta t) = e^{\mat A(t+\delta t)}\vec v(t),
\end{equation}
where the matrix~$\mat A(t+\delta t)$ can be written in terms of univariate integrals, 
\begin{equation}
    \mat A(t+\delta t) = \delta t\mat B_0(t) + (\delta t)^2
    [\mat B_0(t), \mat B_1(t)] 
    + \mathcal O\big((\delta t)^5\big),
\end{equation}
with help of the matrices $ \mat B_j(t)$
\begin{equation}
    \mat B_j(t) = \frac{1}{(\delta t)^{j+1}}
    \int_{-\delta t/2}^{\delta t/2}d\tau \tau^j
    \mat M\left(t+\frac{\delta t}{2} + \tau\right).
\end{equation}
It turns out that it is sufficient to terminate the series after first term, and write the solution as
\begin{equation}
    \vec v(t+\delta t) = e^{\delta t\mat B_0(t)}\vec v(t).
\end{equation}
Now, for the SLED one requires a small time step, and the bottle neck of the above method is the calculation of the matrix exponential. For large and sparse systems, that can be efficiently implemented with the Krylov subspace method. For a small time step~$\delta t$ the matrix~$\mat B_0(t)$ and the vector~$\vec v(t)$ can be accurately expressed in the $m$ dimensional Krylov subspace, 
with $m\ll \dim \mat B_0$. This subspace is spanned by the vectors
\begin{displaymath}
    \Big\{\vec v(t),\, \mat B_0(t)\vec v(t),\, \mat B_0^2(t)\vec v_(t),\,\dots,\, \mat B_0^{m-1}(t)\vec v(t)\Big\}.
\end{displaymath}
Orthonormalizing this subspace results in a unitary matrix
\begin{equation}
    \mat K_m = 
    \begin{pmatrix}
        \vec u_1 & \vec u_2 &\vec u_3 & \dots & \vec u_m
    \end{pmatrix},
\end{equation}
with which one can express the original matrix~$\mat B_0(t)$ as a~$m\times m$ dimensional matrix,
\begin{equation}
    \mat B = \mat K_m^\dag\mat B_0\mat K_m,
\end{equation}
with which one can express the time evolution approximately as
\begin{equation}
    \vec v(t+\delta t) \approx \mat K_m e^{\delta t B}\mat K_m^\dag\vec v(t),
\end{equation}
so that now we only need to calculate the matrix exponential of a small $m\times m$ matrix, instead of the full one. The orthogonalization of the subspace can for non-Hermitian matrix be performed with the Arnoldi iteration, where one first constructs an $(m+1)\times m$ upper Hessenberg matrix $\widetilde {\mat B}$ and $ d\times (m+1)$ dimensional matrix~$\mat K_{m+1}$, where~$d$ is the dimension of original matrix~$\mat B_0$. This can be done with the Gram--Schmidt orthogonalization,
\begin{equation}
    m_{j+1,j}\vec u_{j+1} = \mat B_0\vec u_j - \sum_{i=1}^j
    m_{i,j}\vec u_i,\quad m_{ij} = (\mat B_0\vec u_i)^\dag\vec u_j,
\end{equation}
where~$m_{i,j}$ are the elements of the matrix~$\widetilde{\mat B}$. The desired matrices can then be obtained by discarding the last row of~$\widetilde{\mat B}$ and the last column of~$\mat K_{m+1}$.

\end{document}